\documentclass[final]{article}
\usepackage{neurips_2024}
\setcitestyle{numbers,comma}
\usepackage[utf8]{inputenc} % allow utf-8 input
\usepackage[T1]{fontenc}    % use 8-bit T1 fonts
\usepackage{hyperref}       % hyperlinks
\usepackage{url}            % simple URL typesetting
\usepackage{booktabs}       % professional-quality tables
\usepackage{amsfonts}       % blackboard math symbols
\usepackage{nicefrac}       % compact symbols for 1/2, etc.
\usepackage{microtype}      % microtypography
\usepackage{xcolor}         % colors
\usepackage{setspace}
\usepackage{graphicx}
\usepackage[labelfont=bf]{caption}
\usepackage{caption}
\usepackage{enumitem} 
\title{Simulation of prosthetic vision with the PRIMA system and enhancement of face representation}

\author{
Anna Kochnev Goldstein$^{1*}$, Jungyeon Park$^{2*}$, Yueming Zhuo$^{1}$, Nathan Jensen$^{1}$, Daniel Palanker$^{2,3}$\\[0.5em]
\normalsize
$^1$Department of Electrical Engineering, Stanford University, Stanford, CA, USA\\
$^2$Hansen Experimental Physics Laboratory, Stanford University, Stanford, CA, USA\\
$^3$Department of Ophthalmology, Stanford University, Stanford, CA, USA\\
\textsuperscript{*}These authors contributed equally.
}

\begin{document}

\maketitle
\vspace{4pt}

\begin{abstract}
\textsl{Objective.} Patients implanted with the PRIMA photovoltaic subretinal prosthesis in geographic atrophy report form vision with the average acuity matching the 100$\mu$m pixel size. Although this remarkable outcome enables them to read and write, they report difficulty with perceiving faces. Despite the pixelated stimulation, patients report seeing smooth patterns rather than dots. This paper provides a novel, non-pixelated algorithm for simulating prosthetic vision the way it is experienced by PRIMA patients, compares the algorithm’s predictions to clinical perceptual outcomes, and offers computer vision and machine learning (ML) methods to improve face representation. 

\textsl{Approach.} Our simulation algorithm (ProViSim) integrates a grayscale filter, spatial resolution filter, and contrast filter. This accounts for the limited sampling density of the retinal implant (pixel pitch), as well as the reduced contrast sensitivity of prosthetic vision. Patterns of Landolt C and faces created using this simulator are compared to reports from actual PRIMA users. To recover the facial features lost in prosthetic vision due to limited resolution or contrast, we apply an ML facial landmarking model, as well as contrast-adjusting tone curves to the image prior to its projection onto the photovoltaic retinal implant.

\textsl{Main results.} Prosthetic vision simulated using the above algorithm matches the maximum letter acuity observed in clinical studies, as well as the patients’ subjective descriptions of perceived facial features. Applying the inversed contrast filter to the image prior to its projection onto the implant and accentuating the facial features using an ML facial landmarking model helps preserve the contrast in prosthetic vision, improves emotion recognition and reduces the response time.

\textsl{Significance.} Spatial and contrast constraints of prosthetic vision limit resolvable features and degrade natural images. ML based methods and contrast adjustments prior to image projection onto the implant mitigate some limitations and improve face representation. Even though higher spatial resolution can be expected with implants having smaller pixels, contrast enhancement still remains essential for face recognition.
\end{abstract}


\section{Introduction}
\vspace{-6pt}
Retinal degenerative diseases resulting in the progressive loss of photoreceptors, such as Retinitis Pigmentosa, Stargardt Disease and Age-related Macular Degeneration (AMD), are among the leading causes of incurable blindness today \cite{smith_risk_2001}. While photoreceptors are lost in retinal degeneration, the inner retinal neurons survive to a large extent \cite{mazzoni_retinal_2008, humayun_morphometric_1999, kim_morphometric_2002}. Retinal prostheses are
designed to reintroduce information into the visual system by electrically stimulating these remaining retinal neurons.

PRIMA (Science Corp., Alameda, CA) is a photovoltaic subretinal prosthesis developed to restore vision in the geographic atrophy (GA) region of AMD patients. The 2×2mm PRIMA implant is 30$\mu$ thick and is composed of 378 hexagonally tiled pixels, each 100$\mu$m wide, containing active and return electrodes. Each pixel in this implant converts light into electric current to stimulate the second-order retinal neurons, primarily the bipolar cells \cite{lorach_photovoltaic_2015}. Because photovoltaic pixels require light much brighter than ambient for stimulation and the electric current should be pulsed to preserve the charge balance, images captured by a camera are projected from augmented-reality glasses onto the implant using pulsed intensified light (Figure 1A). To avoid any confounding effects of this intense light on the remaining photoreceptors outside the GA, its wavelength is invisible - in the near-infrared part of the spectrum (880nm). A controller with a pocket computer provides image processing for enhancement of prosthetic vision, while the transparent glasses leave the peripheral natural vision unaffected. Since the photovoltaic implant is powered by light, it has no wires that could affect the surrounding residual photoreceptors.

\begin{figure}[h] % t = top placement
    \centering
    \includegraphics[width=0.6\textwidth, height=11cm]{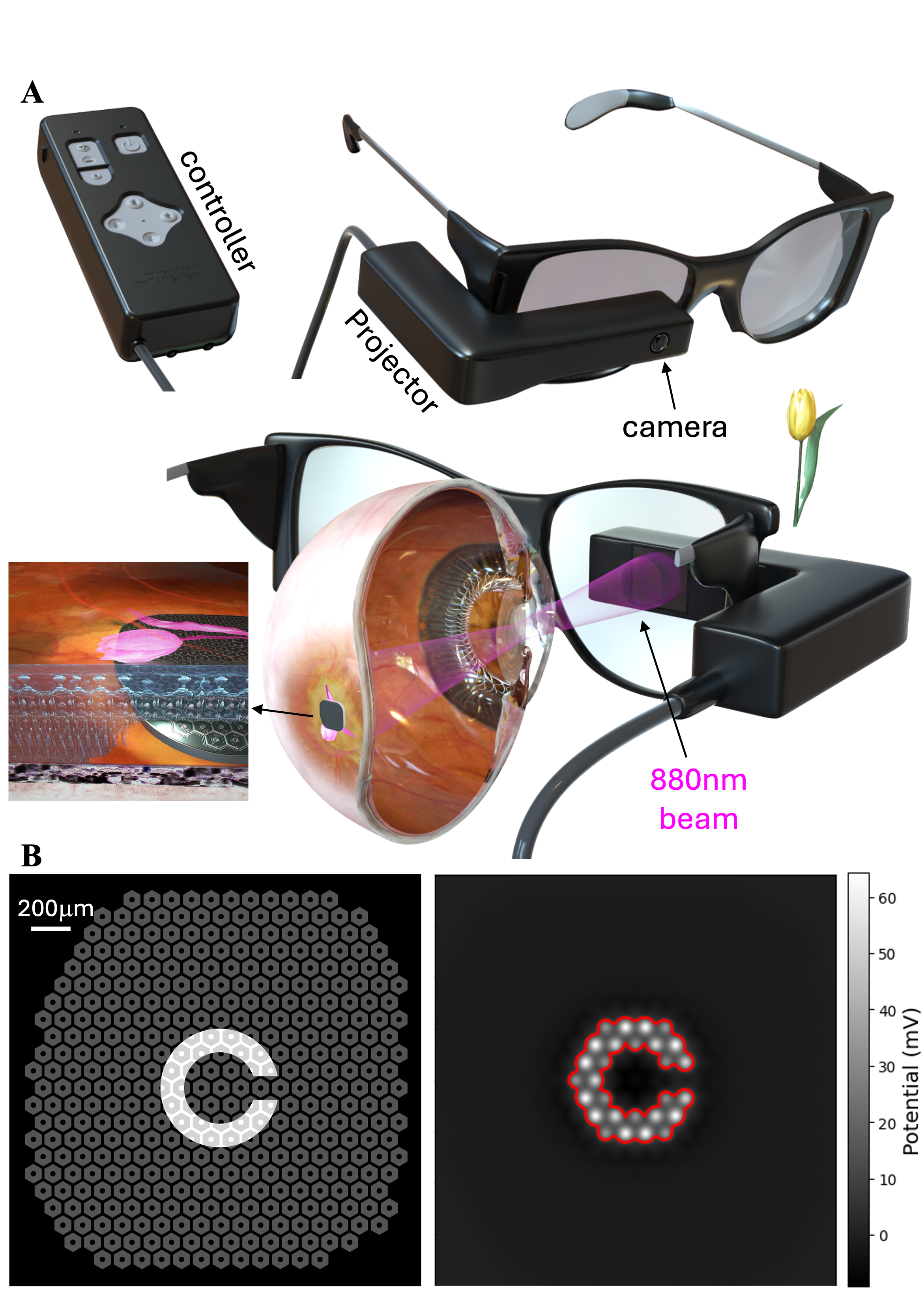} % Adjust width as needed
    \caption{A. A diagram of the PRIMA system, including augmented-reality glasses with a camera, an
            image processor and a projector. Processed images are projected onto the implant using 880nm beam.
            Each pixel in the subretinal implant converts pulsed light into electric current to stimulate the retina. B. Left: A diagram of a photovoltaic array with a projected Landolt C pattern with a gap width of 1.2 pixels. Right: electric potential across bipolar cells in the retina (from 20 to 87µm above the implant). The red contour outlines the stimulation threshold of 11.7 mV.}
    \label{fig:prima}
\end{figure}

In a series of two clinical trials, PRIMA was implanted in 43 AMD patients suffering from central vision loss due to GA. The implant enabled patients to reliably recognize letters and words, with acuity closely matching the maximum spatial resolution allowed by the 100$\mu$m pixels \cite{palanker_photovoltaic_2020, palanker_simultaneous_2022}, which is 20/417. Normal vision of 20/20 corresponds to an angular resolution of 30 cycles per degree, and one degree in a human eye corresponds to 288$\mu$m on the retina \cite{drasdo_non-linear_1974}. Since at least two pixels are required to resolve a cycle, the corresponding equivalent pixel size for normal vision is 288/30/2=4.8$\mu$m. Therefore, 100$\mu$m pixels correspond to acuity about 21 times (100/4.8) worse than normal. Using electronic zoom on the controller, patients improved the resolvable letter size, on average, by 25 ETDRS letters (5 lines) compared to the baseline, demonstrating clinically meaningful benefits of prosthetic central vision \cite{holz_subretinal_2025}. At the same time, patients reported a subjective experience of being able to identify big facial features, such as the hairline, but unable to perceive finer facial details or identify faces. As such, facial perception remains one of their most desired features of prosthetic vision.

To understand whether the facial perception obtained with prosthetic vision can be improved, one needs to envision what patients see when they look at a face using prosthetic vision. Several previous studies have offered illustrations of prosthetic vision, characterizing it as patterns of bright spots in a pixelated image \cite{chen_simulating_2009, wang_sketch_2024, dagnelie_simulations_2011, moneer_enhancing_2023, chang_facial_2012}. However, PRIMA patients describe their precepts as smooth continuous white lines rather than a collection of spots, even though the electric potential generated by the implant is indeed pixelated (Fig. 1B). It is unclear what part of the visual pathway converts the pixelated stimulation pattern applied to the bipolar cells into perception of smooth lines. It may be related to the combination of eye movements with the perceptual completion phenomenon, where we tend to perceive visual features that aren’t present in the scene to reconcile them with familiar or expected patterns and objects \cite{komatsu_neural_2006}. Regardless of the reason, existing simulators that produce pixelated images do not seem to be the right framework to understand prosthetic vision with PRIMA. Other studies suggest substitutes such as simplifying the original detailed visual scene based on the task at hand \cite{beyeler_towards_2022} or modifications such as caricaturing \cite{irons_face_2017}.

This paper offers a novel simulator of prosthetic vision (ProViSim) based on both objective clinical measures and subjective accounts of the PRIMA patients’ experience that aligns with their smooth pattern perception. In addition, based on the results produced by this simulator, it offers a pathway to improve facial perception using computer vision aimed to help patients recognize faces
in the most natural way without relying on scene augmentation. The proposed enhancements, tested on 16 sighted individuals, hold the potential to provide enhanced face perception to existing PRIMA patients via a software update without any changes to the implanted hardware. In addition, we show the expected extent of improvements with the next-generation implants containing
smaller pixels.

\vspace{-2pt}
\section{Simulating the Prosthetic Vision of PRIMA Patients}
\vspace{-3pt}

Retinal degeneration leads to progressive loss of contrast sensitivity and spatial resolution \cite{vingopoulos_measuring_2021}. In prosthetic vision, the density (pitch) of the implant pixels imposes an upper limit on the spatial frequencies that can be sampled and transferred by the implant. Patients with subretinal implants describe their prosthetic vision as monochromatic (variations of white) \cite{palanker_photovoltaic_2020, palanker_simultaneous_2022}, with up to seven distinguishable contrast levels \cite{zrenner_subretinal_2011}. Prosthetic contrast sensitivity in rodents was also found to be about 12\%, corresponding to 8 distinguishable grey levels \cite{ho_temporal_2018}.

Therefore, to simulate prosthetic vision, the algorithm described below converts a color image into grayscale, reduces its resolution to the sampling limit of the implant, and compresses its contrast sensitivity curve. It is important to emphasize that this algorithm does not represent the actual neural signal processing in the human retina or the brain. Instead, it simply reflects the resolution limitations of the hardware as well as the clinical observations of the perceived color and contrast sensitivity reported by the patients.

\subsection{Resolution Reduction}
The PRIMA implant is a 2x2mm photovoltaic array with 100$\mu$m pixels tiled in an
approximate 20x20 grid. Since at least two pixels are required to sample a cycle, the current device imposes a high frequency cutoff of 10 cycles per image. Therefore, we implemented a radial low-pass filter of the Fourier-transformed image using a Tukey window with a radius of 10 cycles
(Figure 2). Repeated sampling of the image with saccadic eye movements helps improve resolution beyond the sampling limit by as much as 30\% \cite{abraham_active_2019, ratnam_fixational_2015, ratnam_benefits_2017}, presumably based on similar principles to the super-resolution algorithms implemented with conventional cameras \cite{ur_improved_1992}. Therefore, to account for the improved acuity associated with microsaccades, we added a 30\% apodization to the low-pass filter, ensuring that frequencies are gradually attenuated beyond the cutoff range with a tapering factor of 0.3.

\begin{figure}[h]
    \centering
    \includegraphics[width=0.9\textwidth]{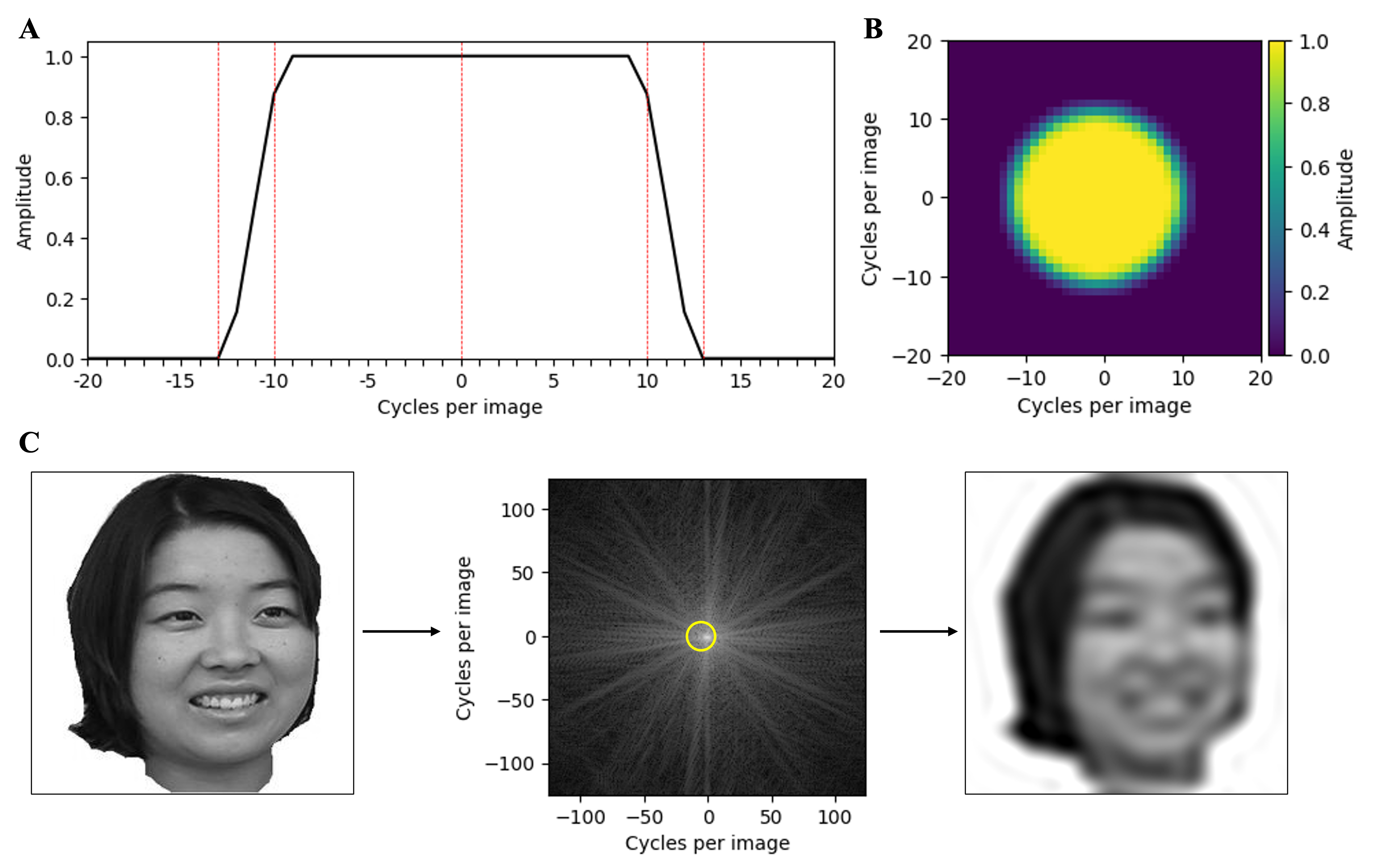} % Adjust width as needed
    \caption{A. Tukey window with a radius of 10 cycles per image and an apodization of 30\%. B. The
            filter’s corresponding mask in the frequency domain. C. The resolution reduction pipeline: an input
            image is transformed into its frequency representation, multiplied by the filter’s mask and transformed
            back to the image domain.}
    \label{fig:spatial_filtering}
\end{figure}

\subsection{Contrast Reduction}
Clinical observations demonstrated the reduced contrast sensitivity in retinal degeneration and in prosthetic vision, but it is unclear whether that loss results in a more uniformly grey or a more distinctly black-and-white perception. Therefore, we simulated contrast sensitivity distortion by two transformations: (a) A Gamma tone curve that converts the original image to a more uniformly grey, and (b) a Sigmoid tone curve that results in a more distinctly black-and-white perception (Figure 3). While the Gamma transform retains normal black (zero brightness) and white (maximum brightness) levels, the Sigmoid transform also includes a horizontal offset parameter that may represent a stimulation threshold and also enables to adjust the black level such that the darkest level is not as dark as normal black.

\begin{figure}[h]
    \centering
    \includegraphics[width=0.9\textwidth]{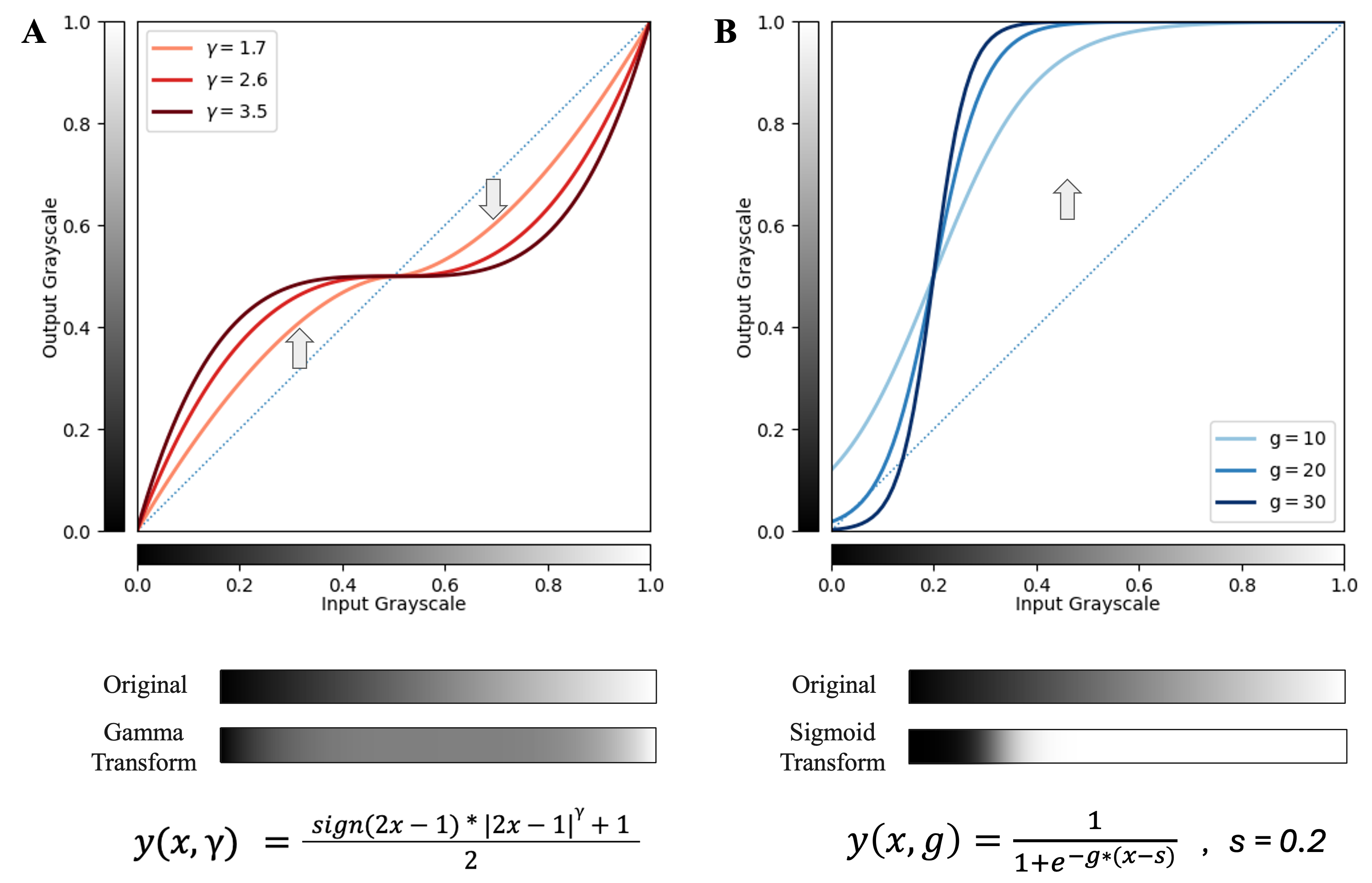} % Adjust width as needed
    \caption{A. Gamma curves along with a sample transform of a grayscale bar and the equation. B. Sigmoid
            curves along with a sample transform of a grayscale bar and the equation for s=0.2.}
    \label{fig:tone_curve}
\end{figure}

Applying these transforms to the Campbell-Robson chart, which visualizes the interaction between spatial resolution and contrast sensitivity (Figure 4A, C), the visible range of the gratings contracts toward the bottom, signifying a loss of contrast sensitivity. The specific gamma (1.7, 2.6, 3.5) and gain (10, 20, 30) values presented here were selected such that the three pairs simulate a similar level of contrast sensitivity loss, with a remaining contrast of approximately 0.1, 0.3 and 0.6. With reduced spatial resolution, the visible range further shrinks to the left (Figure 4B, D).

\begin{figure}[h]
    \centering
    \includegraphics[width=0.9\textwidth]{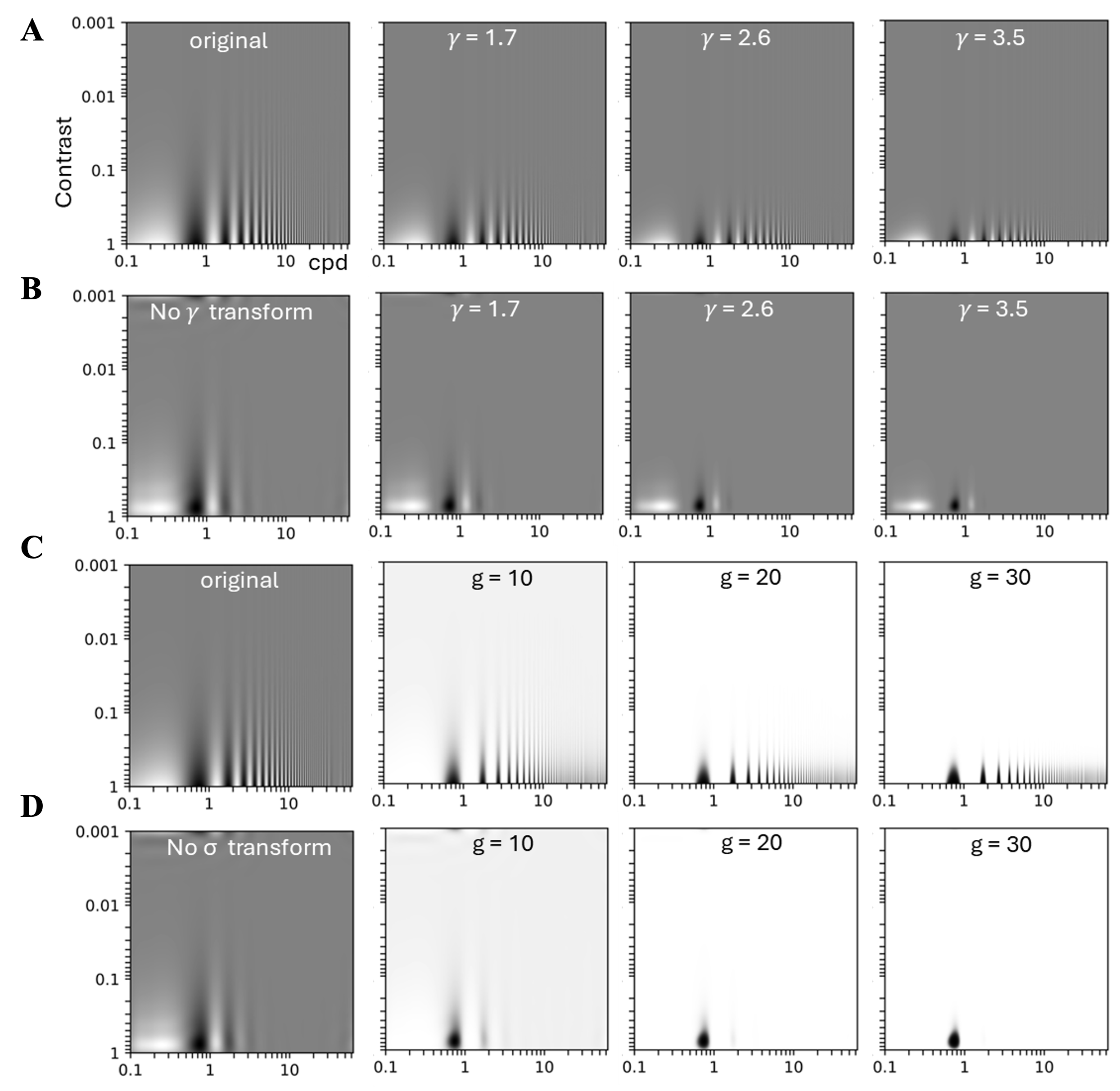} % Adjust width as needed
    \caption{A. Campbell-Robson charts (sinusoidal gratings with variable spatial frequencies (cpd) and
            contrast) at various Gamma transforms. B. The PRIMA-resolution version of the same charts. C. Campbell-
            Robson charts with various Sigmoid transforms. D. The PRIMA-resolution version of the same charts.}
    \label{fig:crc}
\end{figure}

\vspace{-5pt}
\subsection{Testing on Letters}
To ensure that the results of our simulation align with the acuity measurements in the clinical trials, we applied the algorithm to a grid of Landolt C optotypes of several sizes (Figure 5A). In this task, visual acuity is determined by the patient’s ability to identify the orientation of the gap in the letter C. As can be seen in Figure 5C, D, a gap of 0.8 pixels and larger can be resolved with both types of contrast transformations, matching the clinical results in the best PRIMA patients.

\subsection{Simulated Faces}
To simulate the perception of faces in prosthetic vision, the same algorithm of converting the original color image to grayscale, reducing the resolution to the PRIMA sampling limit, and compressing the contrast curve is applied. The simulated outputs are shown in Figure 6. The ringing effect in the output images, particularly visible at the top row, is a computational artifact of the Fourier filtering procedure and is not what patients experience. In principle, this ringing can be suppressed using a more complex filter design. However, since a long-tailed filter that includes higher frequencies cannot be easily justified as accurately representing the physical sampling of the device, we decided to keep this artifact in favor of a more straightforward filtering procedure. As can be seen in Figure 6, using prosthetic vision, one may be able to tell if they are looking at a face and identify the person’s hairline, but not recognize the individual or identify facial expressions or emotions.

\begin{figure}[h]
    \centering
    \includegraphics[width=0.8\textwidth]{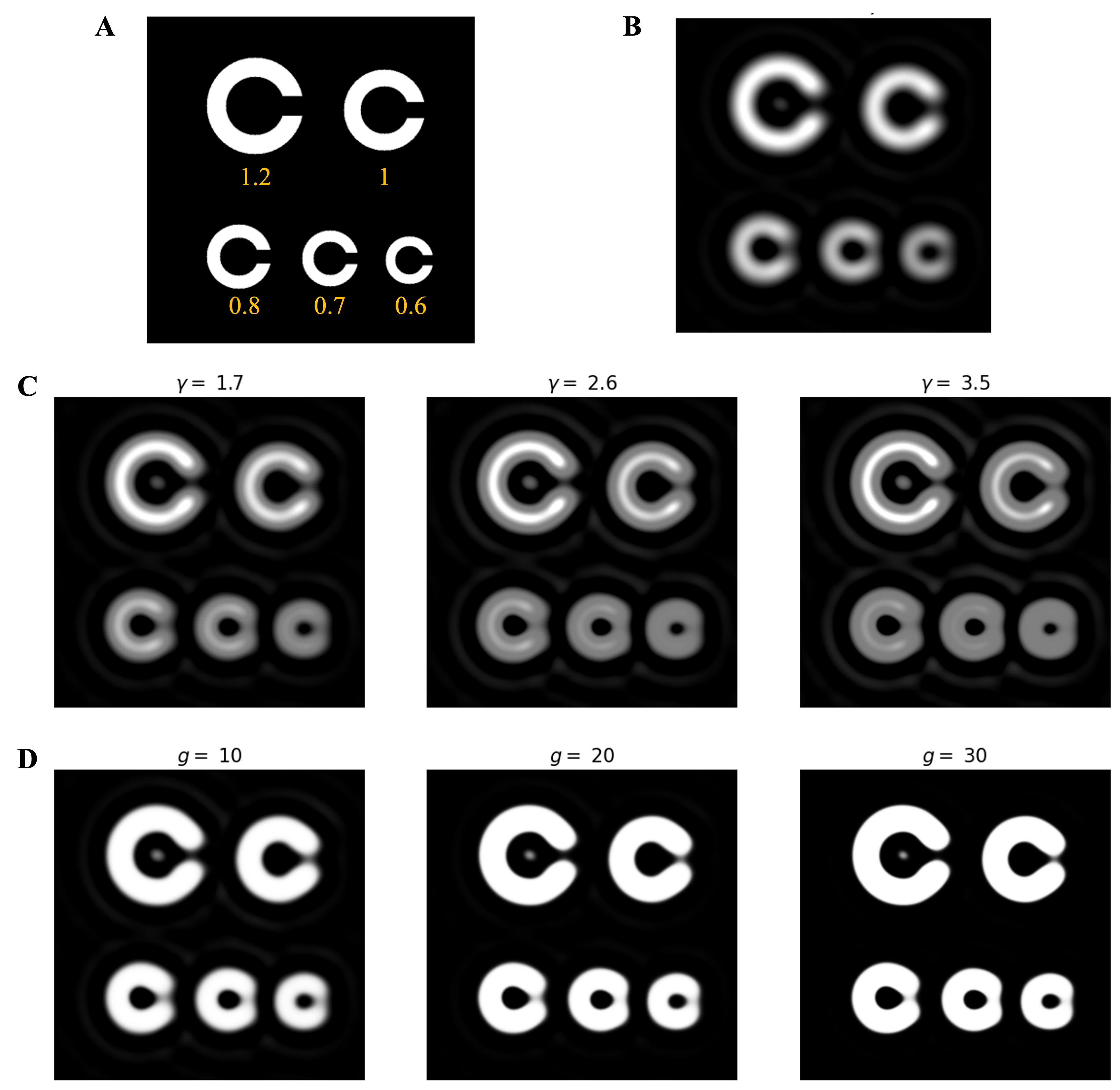} % Adjust width as needed
    \caption{A. Landolt C letters with gap sizes varying from 1.2 to 0.6 pixels in width. The black square
            represents the 20x20 pixel grid size. B. The same letters with the resolution reduction associated with the PRIMA implant. C. The same letters for various Gamma transforms. D. The same letters for various Sigmoid transforms. }
    \label{fig:landolt_c}
\end{figure}

Most PRIMA patients did not look at faces as this wasn’t the focus of the trial and they were not directly instructed to do so. However, three patients that did look at faces described them as “cloud white” or “completely blurred” and reported that they could partially see where the face ends and the background begin, the outline of the face, and the presence of the hair and nose. One patient could also recognize the hair outline, hair style, facing direction, and even partially see the eyes and lips. Although it remains unclear which one of the contrast reduction transformations is the most accurate for prosthetic vision with PRIMA, the illustrations in the two
right columns of the sigmoid transform stand in good agreement with these subjective descriptions of face perception obtained from the patients.

\section{Enhancing the Perception of Facial Features}
\subsection{Improving Contrast Sensitivity}
If the contrast reduction function of prosthetic vision is known, we could minimize its effect by applying the inverse function before projecting the image into the eye. Figure 7 shows the results of applying the inverse of the Gamma and the Sigmoid curves, respectively, to the original image before passing the resulting image through our prosthetic vision simulator. Comparison of the “original” columns to the “inverse” ones demonstrates that transforming the contrast of the original image in a way that counters the loss caused by prosthetic vision restores some additional levels of gray. This allows better representing some important facial planes, such as the eye sockets, forehead, and cheek bones.

\begin{figure}[htbp]
    \centering
    \includegraphics[width=1.0\textwidth]{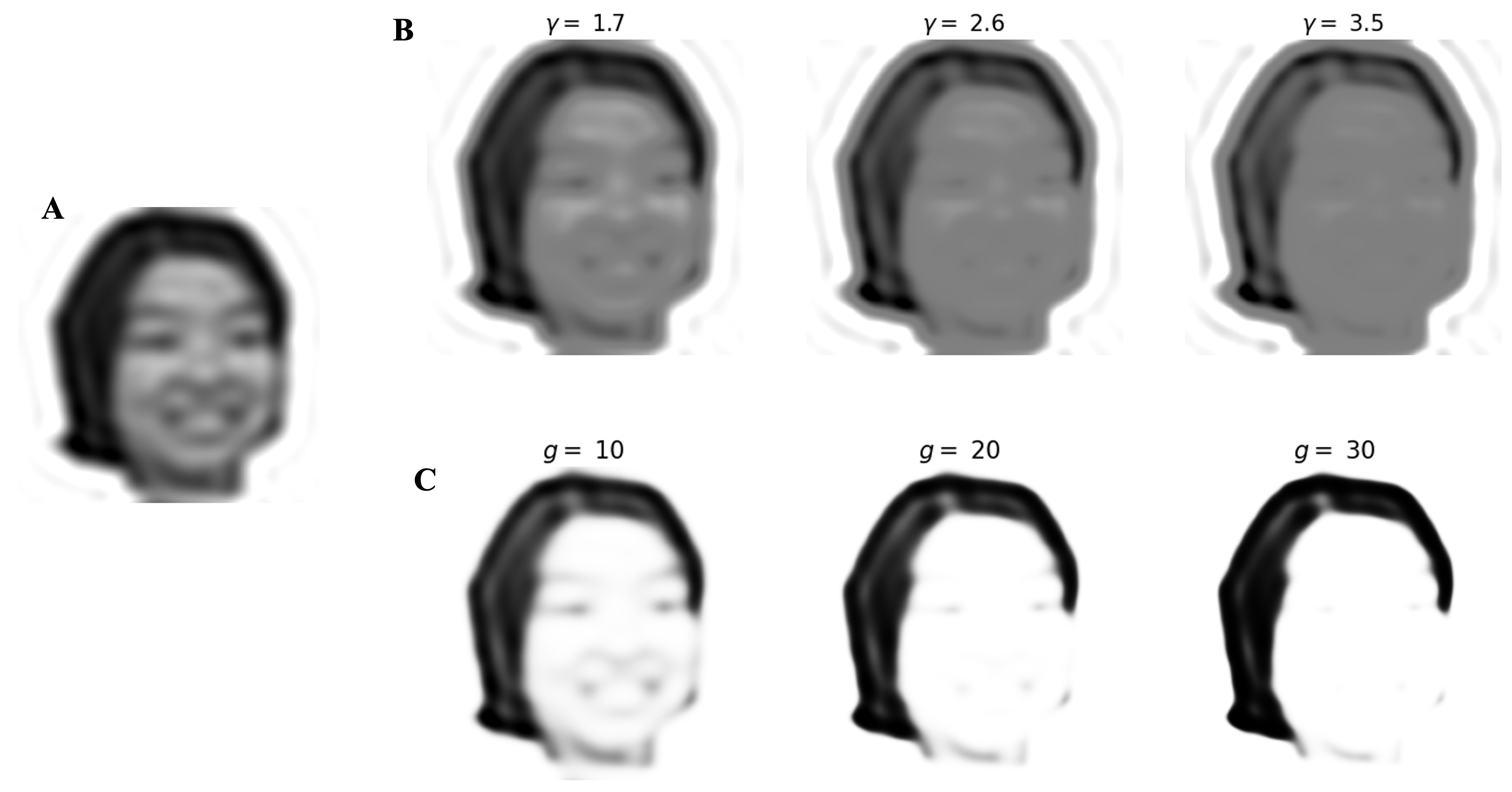} % Adjust width as needed
    \caption{A. An image of a face with spatial resolution reduced to the PRIMA sampling limit. B. The same
            image with various levels of contrast reduction using the Gamma transform. C. The same image with
            various levels of contrast reduction using the Sigmoid transform.}
    \label{fig:face_provisim}
\end{figure}

One limitation in such an inversion is the fact that the DMD projector in the PRIMA glasses provides only 14 discrete levels of stimulation (adjusted by the micromirrors’ ON time from 0.7 to 9.8ms in 0.7ms increments), which limits the fidelity of any "inversed" transformation. However, this hardware limitation might be alleviated in the future. More importantly, since in our simulations, we know the exact function used to simulate the loss of contrast sensitivity, it is straightforward to apply the exact inverse of that function prior to the simulation algorithm and get the result shown in Figure 7. However, since we don’t know yet the exact function governing the contrast loss in PRIMA patients, clinical testing is required to assess the contrast loss in individual patients before applying an inverse function.

\begin{figure}[h]
    \centering
    \includegraphics[width=1.0\textwidth]{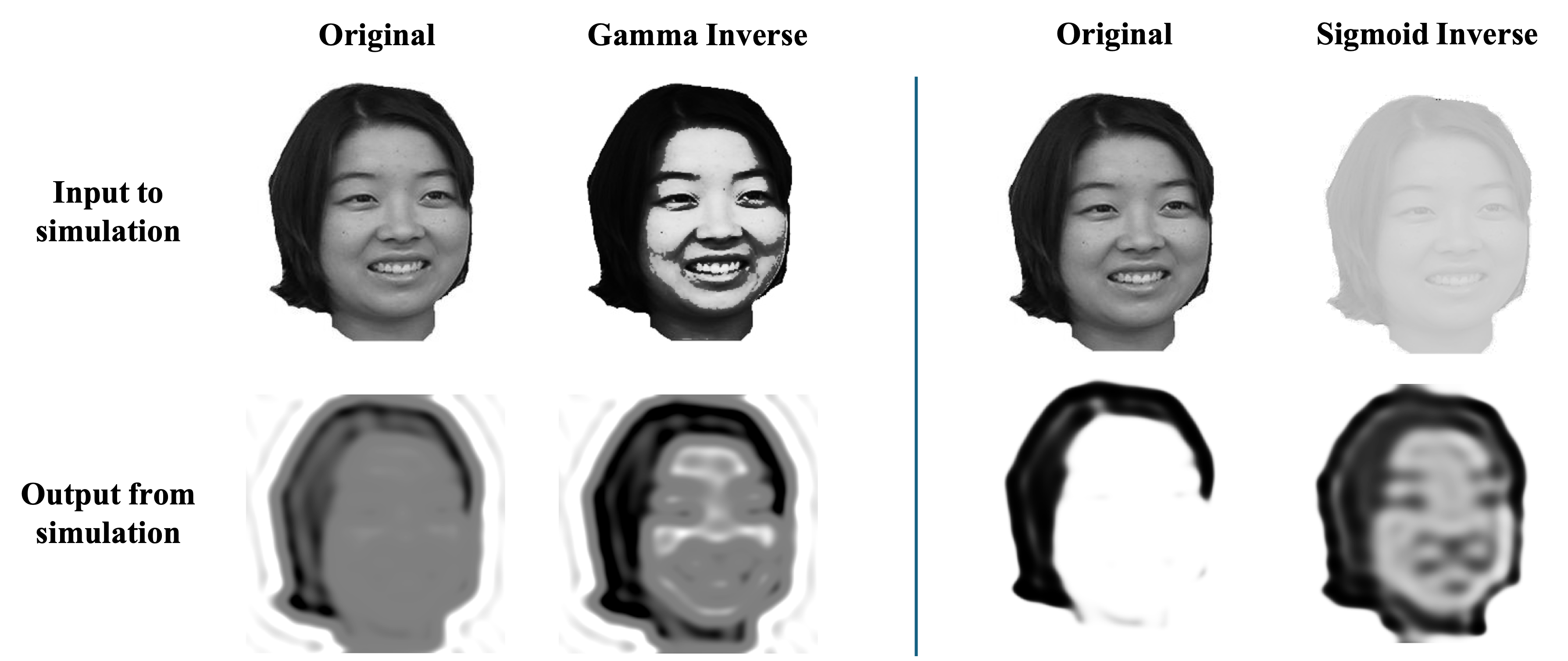} % Adjust width as needed
    \caption{Simulated face perception with and without preemptive contrast adjustment with a Gamma
            inverse tone curve (left) and a Sigmoid inverse tone curve (right).}
    \label{fig:inverse_tone_curve}
\end{figure}

\subsection{Reintroducing Facial Features}
Spatial and contrast filters of simulated prosthetic vision with PRIMA remove most of the finer facial details, leaving just the face outline and hair. We can improve this outcome by making the fine facial features more pronounced in color or thickness in the original image before applying the simulation algorithm. To achieve this, we first identify the facial landmarks using the open-source, real-time “MediaPipe Face Landmarker” machine learning model from Google on the original color image. Then, we can thicken and darken the contours of the obtained key facial features - eyebrows, irises, and lips - to enable their passage through the filters. To a normally sighted individual, the bolder features appear unnatural, but after the filters of prosthetic vision, the key facial elements reappear attenuated and closer to natural size. The results of this procedure are shown in Figure 8.

\vspace{-2pt}
\begin{figure}[h]
    \centering
    \includegraphics[width=0.9\textwidth]{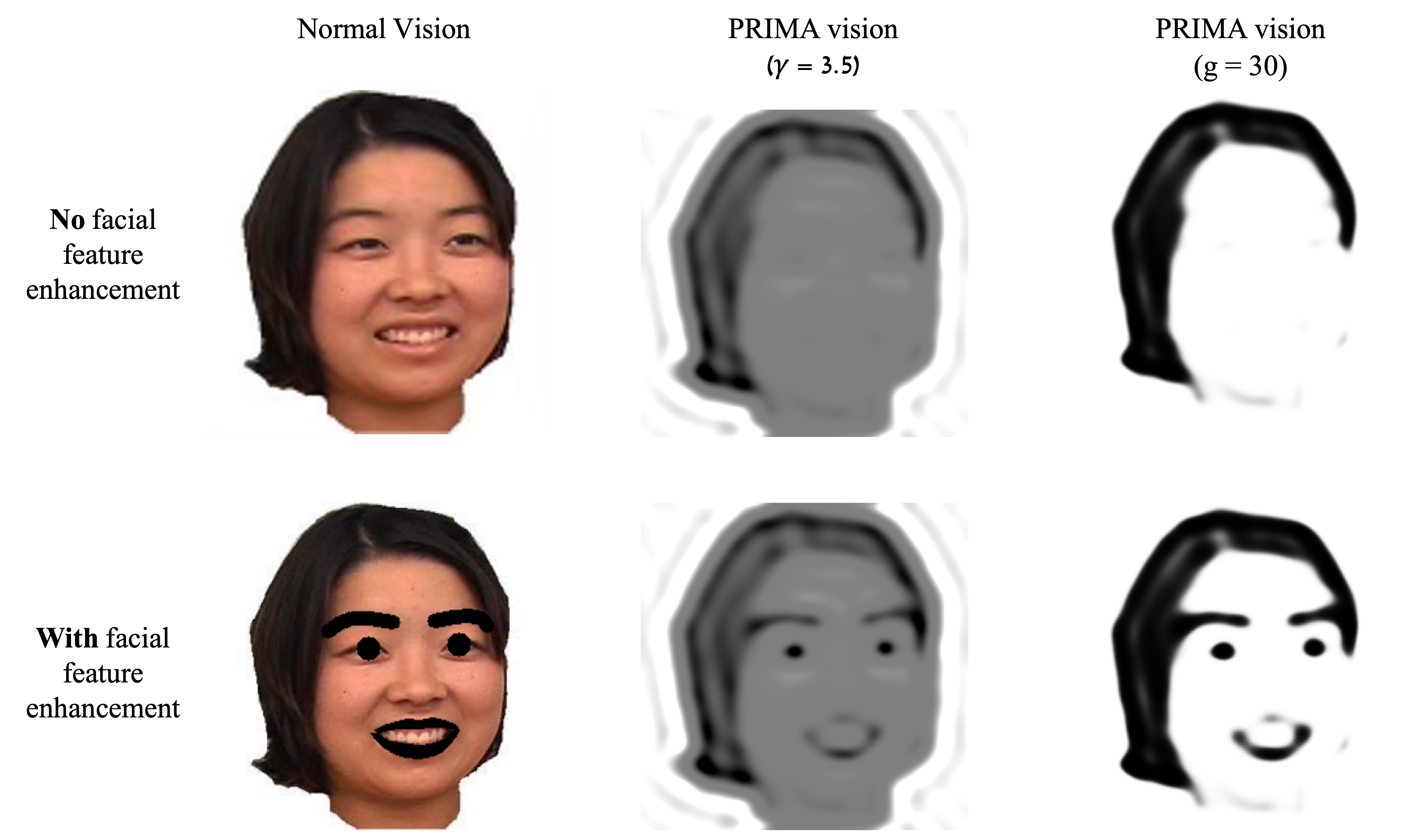} % Adjust width as needed
    \caption{Simulated prosthetic vision without (top line) and with (bottom line) landmarking-based facial
            feature enhancement. The landmarking used is 0.7-implant-pixels thick and black in color.}
    \label{fig:facial_landmarking}
\end{figure}

Figure 9 shows three other people expressing different emotions – disgust, surprise, and
happiness, enhanced with landmarking, as well as landmarking and contrast combined. Without
any enhancements, one can hardly identify any of the emotions. However, with preemptive facial
landmarking, and even more so with both landmarking and contrast correction, the emotions
become much easier to identify and the direction of gaze more pronounced. This should enable the
patients to perceive a smile and establish eye contact - critical aspects of human communication
and connection. As this enhancement is purely computational and can run in real-time at video rate
(30 fps), it can be provided to existing patients as a software update without requiring any hardware
modifications.

\begin{figure}[h]
    \centering
    \includegraphics[width=1\textwidth]{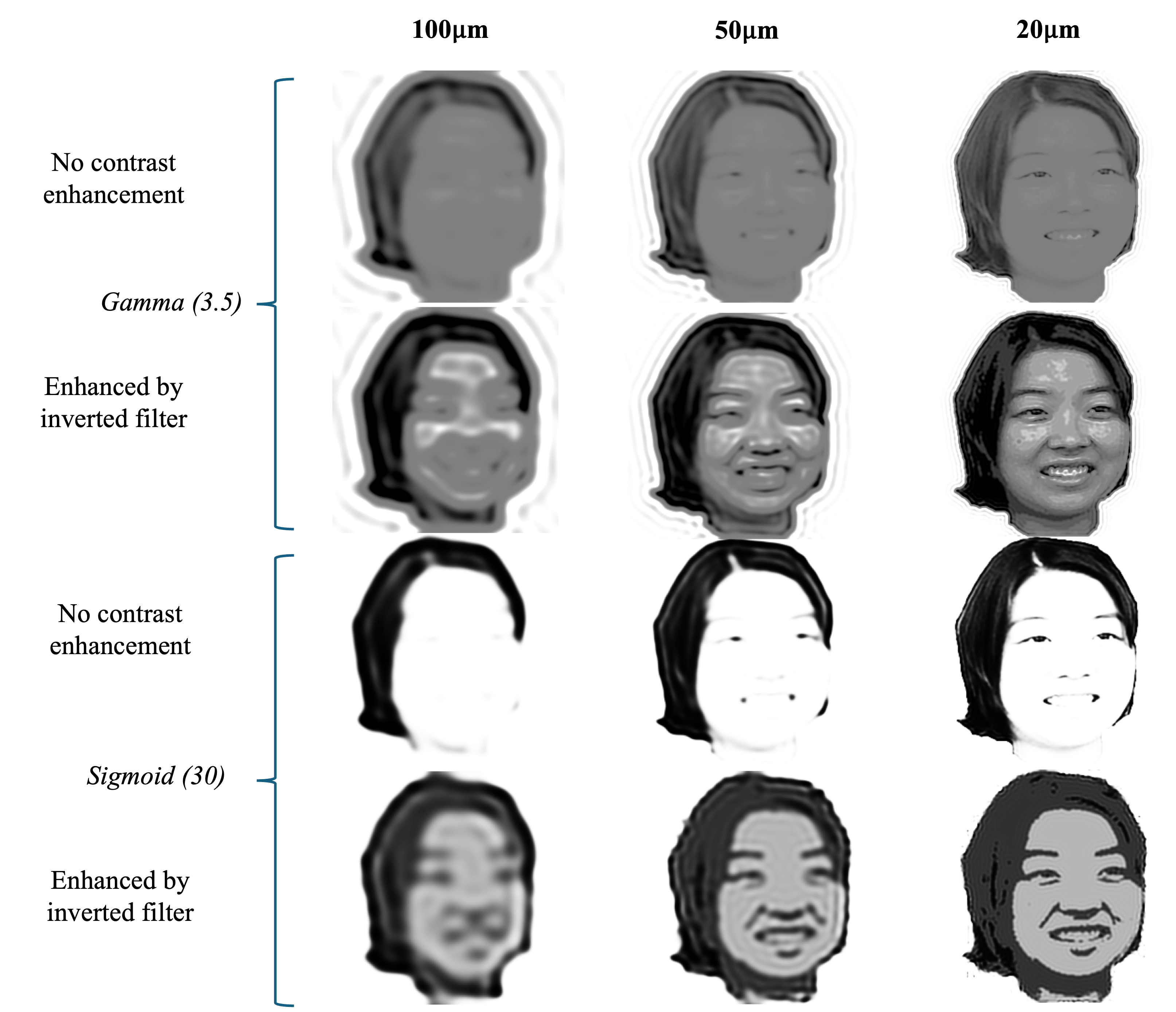} % Adjust width as needed
    \caption{Simulated prosthetic vision of people expressing different emotions – disgust, surprise, and
            happiness, without enhancements (second column), with landmarking (third column), and with both
            landmarking and contrast enhancements (fourth column). In the third column, the landmarking is 0.7-
            implant-pixels thick in dark grey. In the fourth column, the landmarking is 0.3-implant-pixels thick and 50\% darker than the natural color of the feature, for a softer effect.}
    \label{fig:emotions}
\end{figure}

Figure 9 also illustrates the difference between light and dark skin tones. While lighter skin tones lose most inner facial features when processed by the simulator, darker skin tones maintain much more useful information. In fact, they can benefit from slightly accentuating the facial features with facial landmarking, but applying contrast correction makes them unresolvedly dark.
Therefore, no contrast correction was applied on the bottom row. Darker hair colors also remain more visible and help obtain the outline of the face and separate it from the background. The relative contrast between the face and the hair is also significant. 

As contrast correction already restores a lot of the information in faces with lighter skin tones, the landmarking applied on these images was much gentler in both thickness and color. In practice, the thickness of the drawn features, their color, and the application of contrast correction
can be determined in real-time based on the specific face being viewed and the individual preferences of each patient.

\subsection{Face Perception with Future Implants }
The 100$\mu$m pixels in PRIMA impose a hard limit on the spatial frequencies that can be perceived by patients. To improve the resolution limit, a new chip with smaller pixels has been tested in rats and has shown to provide higher resolution \cite{wang_electronic_2022}. This technology is currently being transferred to Science Corporation for the fabrication of its human version.
To illustrate the potential benefits of smaller pixels for face perception, we simulated prosthetic vision with 2x2mm implants composed of 50$\mu$m and 20$\mu$m pixels, with corresponding cutoff frequencies of 20 and 50 cycles per image. 

As shown in Figure 10, smaller pixels enable significantly higher spatial frequencies, allowing the finer details of the human face, such as the eyes, nose, eyebrows and lips, to be sampled by the device and visualized. However, despite the improvement in spatial resolution, the reduced contrast of prosthetic vision still makes the simulated image harder to perceive compared to the original. Applying inversed contrast filters and facial landmarking prior to the simulation algorithm optimizes the advantage of smaller pixels.
Implants with much higher resolution, such as 20$\mu$m pixels, might require only the inverse contrast filters without facial landmarking, but that will be determined based on individual patients’ preference in the future clinical trials.

\begin{figure}[h]
    \centering
    \includegraphics[width=1\textwidth]{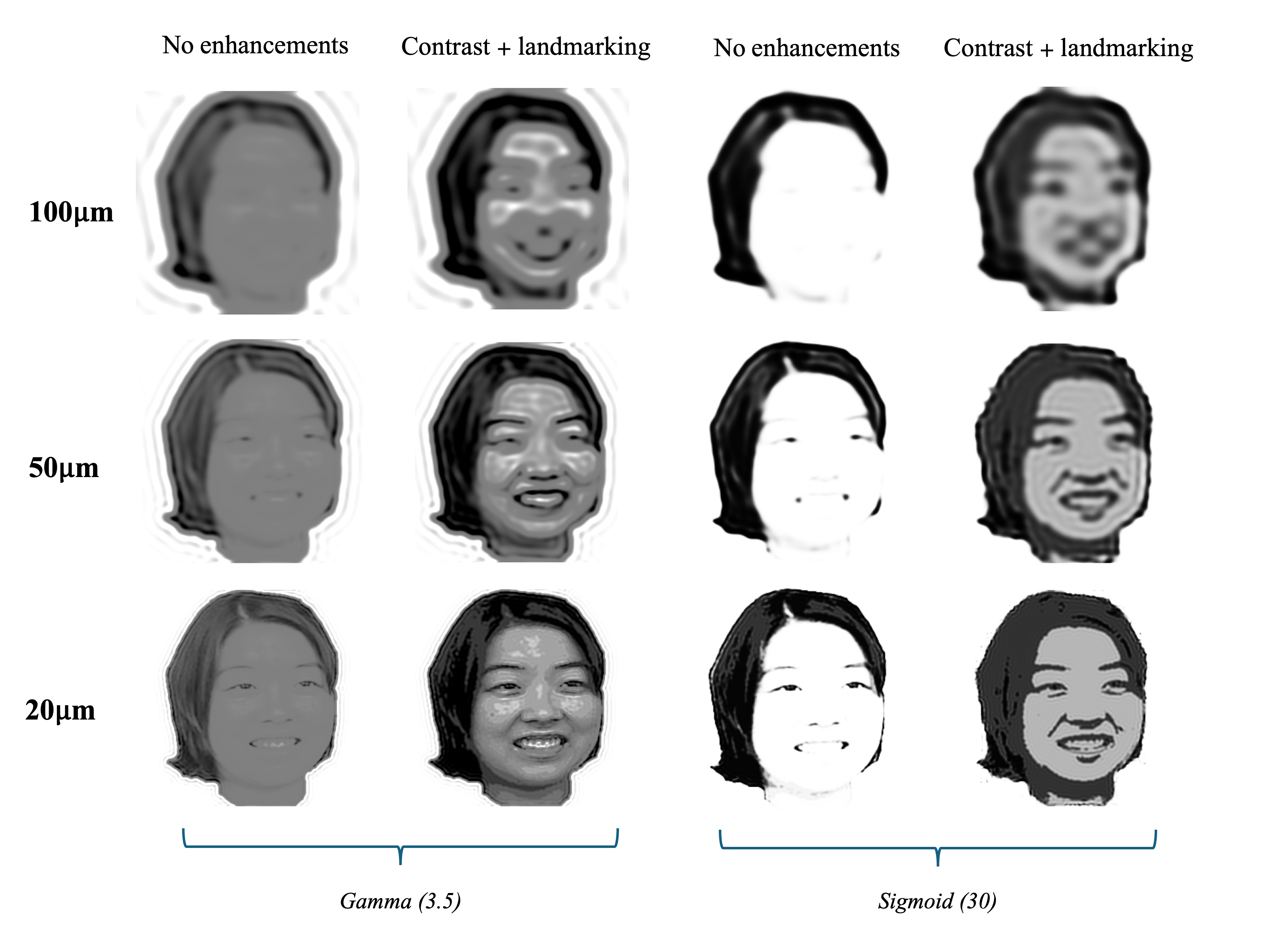} % Adjust width as needed
    \caption{Simulated prosthetic vision of a face for different pixel sizes without and with the enhancements.
            From top to bottom: 100$\mu$m-wide pixels corresponding to PRIMA, 50$\mu$m-wide pixels and 20$\mu$m-wide pixels. The left two columns represent the Gamma (3.5) contrast transform. The right two columns –
            the Sigmoid (30) transform.}
    \label{fig:smaller_implants}
\end{figure}

\subsection{Testing Face and Emotion Recognition on Sighted Individuals}
To test the extent of face perception with prosthetic vision and the efficacy of the proposed
enhancements, we conducted a study on 16 sighted individuals (9 males and 7 females) drawn
from the student body of Stanford University. The participants were positioned in front of a
computer screen displaying a question with four faces as the possible answers. For each question,
the participants were requested to choose the correct image. The questions were: (1) Which person
is the odd one out? (2) Which person is a different gender? and (3) Which face looks happy, sad,
surprised, disgusted, angry, confused, fearful, or neutral.

The trial consisted of two phases: first, the images were processed to simulate prosthetic
vision with PRIMA without any enhancements, whereas in the second phase, the same images
were enhanced by an inversed tone curve and facial landmarking before applying the prosthetic
vision flow (similar to the illustrations in Figure 9). Each type of question was repeated 24 times
throughout each phase with varying faces. As there were 8 emotions and 24 repetitions, each
emotion was displayed 3 times during each phase. Consequently, the accuracy of any given
emotion within each phase was either 0, 33.3, 66.6, or 100\%. The tone curve applied in both cases
to simulate PRIMA vision was the Sigmoid transform, as it seemed to be in better agreement with
the descriptions of the patients. A gain of 30 and a shift of 0.2 were chosen to simulate profound
contrast sensitivity loss (as illustrated in Figure 6C). The participants had 20 seconds to choose
their answer on each screen, and the response time was recorded. Additional 5 participants that
didn’t take part in the original trial were recorded performing the same protocol but using the
natural, unprocessed images, to obtain a baseline of the average response time for each task.

The easiest task was identifying the odd person out. Even in the first phase, relying on the
hair and the general outline of the face, participants demonstrated an average accuracy of 96.1\%, with no significant increase in the second phase. The average response time decreased from 5.12
to 4.32 seconds (p<0.001), with the baseline response time for this task being 3.33 seconds. Gender
was also identified with a relatively high average accuracy of 76.8\% in the first phase, with no
significant increase in the second phase. The response time decreased from 6.52 to 5.84 seconds
(p<0.05), with the baseline for this task being 4.17 seconds.

As expected, the most challenging task was recognizing emotions. During the first phase,
the average accuracy was 25.5\% - equivalent to a random choice between four images, indicating
the lack of internal facial features essential for emotion recognition. During the second phase, the
accuracy increased to 48.4\% (p<0.001), demonstrating the benefits of the image enhancements for
this task. In addition, the average response time dropped from 8.17 to 6.45 seconds (p<0.001),
close to the baseline response time for this task, which was 6 seconds. This outcome is even more
significant since some of the presented emotions, such as sad, neutral, and disgusted, were very
nuanced. For more obvious emotions, the accuracy of “happy” increased from 35.4\% to 83.3\%
(p<0.001), “confused” - from 25\% to 56.25\% (p<0.01), “fearful” - from zero to 56.2\% (p<0.001),
“surprised” – from 37.5\% to 64.6\% (p<0.05), and “angry” from 35.4\% to 56.2\% (p<0.05).

\section{Discussion}
In this paper, we presented an algorithm that simulates the loss of color, resolution and
contrast in prosthetic vision, as described by PRIMA patients - ProViSim. Processing images with
this simulator illustrates why PRIMA patients can resolve letters but cannot perceive faces: unlike
black-and-white letters, simulated prosthetic vision does not reproduce fine facial features, such
as the eyebrows and lips, or the complex greyscale of the various facial planes.

ML-based facial landmarking helps identify the location and outline of the thinner facial
features and make them programmatically more pronounced for any given image in real time. This
procedure ensures that such features remain visible even in a low-resolution and low-contrast
image of prosthetic vision. Similarly, understanding the specific characteristics of contrast
sensitivity loss experienced by patients will enable the application of the inverse of this function
to the image before it is projected into the eye, thereby better retaining information on the facial
structure.

Applying these enhancements on sighted individuals has significantly reduced their
response time and improved their ability to recognize facial expressions and the underlying
emotions. Moreover, if the function that governs contrast sensitivity loss is known, the color of the
landmarking can be chosen such that it creates the thinnest and most natural overall output when
perceived with the limitations imposed by prosthetic vision. This will require a personalized
clinical investigation of a patient’s contrast sensitivity loss function and adjustment of their
preferred enhancement settings.

For reading tasks, PRIMA patients have expressed a preference of viewing white letters on
a black background, which effectively minimizes the percentage of the illuminated area in their
prosthetic field of view. However, the same preference is unlikely to work for faces because they
consist of several different planes and we are used to associate darkness with depth.

The findings of this study and the remaining open questions discussed here will be used as
a foundation for a series of follow-up tests on PRIMA patients intended to measure their individual
contrast sensitivity curves and examine their face perception with and without the suggested
enhancements. Integrating these or similar software enhancements into the PRIMA system should
enable current patients to obtain their most wanted “upgrade” of face perception without any
hardware changes or surgical intervention. In addition, even though the next-generation implants
with smaller pixels should allow the representation of much higher spatial frequencies, they will
still require additional image processing for contrast enhancement in order to optimize the face
perception.

\vspace{10pt}
\subsection*{Declarations}
\vspace{10pt}

\subsection*{Ethics approval and consent to participate}
\vspace{-5pt}
This study was approved by the Stanford IRB panel on human subject research. Sixteen subjects recruited from Stanford University provided consent and participated in the study.

\subsection*{Consent for publication}
\vspace{-5pt}
Not applicable.

\subsection*{Availability of data and materials}
\vspace{-5pt}
\noindent The algorithm is implemented in Python and the code is openly available at \url{https://doi.org/10.5281/zenodo.14921121}.

\vspace{5pt}
\noindent Face images courtesy of Michael J. Tarr, Carnegie Mellon University, \url{http://www.tarrlab.org/}. Funding provided by NSF award 0339122.

\subsection*{Competing interests}
\vspace{-5pt}
\noindent Daniel Palanker: C (consultant) and P (patents licensed by Stanford University to Science Corp).
\noindent All other authors have no conflicting interests to declare.

\subsection*{Funding}
\vspace{-5pt}
\noindent Studies were supported by the National Institutes of Health (Grants R01-EY-035227, and P30-EY-026877), the Department of Defense (Grant W81XWH-22-1-0933), AFOSR (Grant FA9550-19-1-0402), and unrestricted grant from the Research to Prevent Blindness.

\subsection*{Authors' contributions}
\vspace{-5pt}
\noindent AKG, JP, and DP conceived the study. AKG, JP, YZ, and DP developed the methodology. JP and AKG implemented the computational framework. AKG collected the data. AKG and JP wrote the manuscript. JP, AKG and NJ prepared the figures. All authors reviewed and edited the manuscript. 

\newpage

\bibliographystyle{unsrtnat}
\bibliography{References}

\end{document}